\documentclass[a4paper,10pt]{article}

\usepackage{amsmath,amssymb,mathtools}     
\usepackage{color}
\usepackage{graphicx}
\usepackage{subfigure}
\usepackage{cite}                
\usepackage{hyperref}            
\usepackage{multirow,makecell}   
\usepackage{textcomp}
\usepackage{wasysym}

\usepackage[text={17.2cm,25.2cm},centering]{geometry}

\numberwithin{equation}{section}   

\usepackage{float}
\usepackage{subfloat}
\usepackage{amsfonts, mathrsfs}
\usepackage{latexsym}
\usepackage{pdfsync}
\usepackage{marvosym}
\usepackage{diagbox}
\usepackage{epstopdf}

\numberwithin{equation}{section}

\def \be {\begin{equation}}
\def \ee {\end{equation}}
\def \ba {\begin{array}}
\def \ea {\end{array}}
\def \bea{\begin{eqnarray}}
\def \eea{\end{eqnarray}}
\def \nn {\nonumber}

\def \a {\alpha}
\def \b {\beta}
\def \g {\gamma}

\def \e {\epsilon}

\def \m {\mu}
\def \n {\nu}
\def \k {\kappa}

\def \S {\Sigma}
\def \r {\rho}

\def \p {\partial}
\def \f {\frac}

\def \sr {\sqrt}

\def \inf {\infty}

\def \tr {\textrm{tr}}

\def \and {{\textrm{and}}}

\begin{document}

\title{The entanglement properties of holographic QCD model with a critical end point}
\author{Zhibin Li$^1$\footnote{lizhibin@zzu.edu.cn}\,
	Kun Xu$^{2, 3}$\footnote{xukun@ihep.ac.cn}\,
	and Mei Huang$^2$\footnote{huangmei@ucas.ac.cn}}
\date{}

\maketitle

\vspace{-10mm}

\begin{center}
	{\it
		$^{1}$ School of Physics, Zhengzhou University,\\
		No.100 Science Avenue, Zhengzhou 450001, P.R.\,China\\ \vspace{1mm}
		
		$^{2}$ School of Nuclear Science and Technology, University of Chinese Academy of Sciences, Beijing 100049, China\\ \vspace{1mm}
		
		$^{3}$ Institute of High Energy Physics, Chinese Academy of Sciences, Beijing 100049, P.R. China
	}
	\vspace{10mm}
\end{center}

\begin{abstract}
We investigate different entanglement properties of a holographic QCD (hQCD) model with a critical end point at finite baryon density. Firstly we consider the holographic entanglement entropy (HEE) of this hQCD model in a spherical shaped region and a strip shaped region, respectively, and find that the HEE of this hQCD model in both regions can reflect QCD phase transition. What is more is that although the area formulas and minimal area equations of the two regions are quite different, the HEE have very similar behavior on the QCD phase diagram. So we argue that the behavior of HEE on the QCD phase diagram is independent of the shape of subregions. However, as we know that HEE is not a good
quantity to characterize the entanglement between different subregions of a thermal system. So we then study the mutual information (MI), conditional mutual information (CMI) and the entanglement of purification (Ep) in different strip shaped regions. We find that the three entanglement quantities have very similar behavior:  their values do not change so much in the hadronic matter phase and then rise up quickly with the increase of $T$ and $\m$ in the QGP phase. Near the phase boundary, these three entanglement quantities change smoothly in the crossover region, continuously but not smoothly at CEP and show discontinuity behavior in the first phase transition region. And all of them can be used to distinguish different phases of strongly coupled matter.
\end{abstract}

\baselineskip 18pt
\thispagestyle{empty}
\newpage

\tableofcontents

\section{Introduction}
\label{sec-int}

Entanglement property plays a very important role in strongly coupled system. In a quantum many body system, entanglement entropy is a measurement of quantum correlation between different parts of the system \cite{VonNeumann1932}. With AdS/CFT correspondence or more general Gauge/Gravity duality \cite{Maldacena:1997re,Gubser:1998bc,Witten:1998qj,Aharony:1999ti}, the holographic entanglement entropy opens a window to quantum information and quantum gravity \cite{VanRaamsdonk:2010pw,Faulkner:2013ica,Swingle:2014uza,Freedman:2016zud,Czech:2017ryf}.

 From the quantum field theory side, e.g., quantum chromodynamics (QCD), the whole system we considered is on a 4-dimensional Minkowski spacetime and for any state at a fixed time $t_0$ we have the state vector $|\Psi(t_0) \rangle$ and the density matrix
\be
\r=|\Psi(t_0)\rangle\langle\Psi(t_0)|.
\ee
To investigate the entanglement entropy between different parts of this system we firstly divide the whole time slice into two parts which we denote as $A$ and $\bar{A}$ where $A$ is a subregion of the time slice and $\bar{A}$ its complement.  Then we get the reduced density matrix of subsystem $A$ by tracing out the degree of freedom of subsystem $\bar{A}$ in the Hilbert space
\be
\r_A=\tr_{\bar{A}} \r.
\ee
The entanglement entropy of subsystem $A$ can be defined as the von Neumann entropy \cite{VonNeumann1932}
\be
S_A=-\tr(\r_A \log\r_A).
\ee
However, it is not easy to calculate the entanglement entropy directly in the QCD side by using this formula. While according to the AdS/CFT correspondence or AdS/QCD correspondence \cite{Aharony:1999ti,Erdmenger:2007cm,deTeramond:2012rt,Kim:2012ey,Adams:2012th} we know that the holographic duality of entanglement entropy between boundary region $A$ and its complement is the holographic entanglement entropy which can be calculated by the Ryu-Takayanagi formula \cite{Ryu:2006bv,Ryu:2006ef}
\be
S_A\equiv S_A^h=\f{Area(\min_{m(A)\sim A}\{m(A)\})}{4G_N}=\f{Area(\g_A)}{4G_N}=\f{2\pi }{\k^2}Area(\g_A),
\ee
where $m(A)$ is a 3-dimensional surface in the bulk which is homologous to $A$. And the holographic entanglement entropy equals to the area of minimal $m(A)$ which is denoted as $\g_A$ (the R-T surface) over a constant $4G_N$. There have been some efforts to investigate the relation between the holographic entanglement entropy and phase transition in holographic QCD models \cite{Zhang:2016rcm,Knaute:2017lll,Ali-Akbari:2017vtb}. Like in \cite{Zhang:2016rcm} the behavior of holographic entanglement entropy along temperature at zero baryon chemical potential has been shown, and in \cite{Knaute:2017lll} the authors studied the holographic entanglement entropy in a strip shaped region for a holographic QCD model. They both found that holographic entanglement entropy is sensitive to the phase transition of QCD matter.

Another very important part of holographic entanglement entropy is the shape dependence of the subregion $A$ \cite{Dong:2016wcf,Bianchi:2016xvf,Cavini:2019wyb}. Because QCD theory lives on a 4-dimensional spacetime, it will be very difficult to calculate the holographic entanglement entropy for a general shaped region $A$. So we just consider two different shapes in our work to study the shape dependence of region $A$, one is a spherical shaped region and another is a strip shaped region.

For a thermal system with finite temperature $T$ and chemical potential $\m$, entanglement entropy is not a good quantity to measure the entanglement between different subsystems because of the contributions from the thermodynamics. To understand this point more exactly firstly let' s consider the purification of the quantum state on the boundary of a Schwarzschild-AdS black hole. It has been shown that the purified state lives on the double boundary of the bulk spacetime which can be denote as $B$ and $\bar{B}$. If we divide $B$ to be two disjoint subregions $A$ and $\bar{A}$ then the holographic entanglement entropy of subregion $A$ measures the entanglement between $A$ and its complement $\bar{A}\cup \bar{B}$ ( but not $\bar{A}$ )\cite{Chen:2017ahf}. By using the subadditivity and strong subadditivity of entanglement entropy one can define two nonnegative entanglement quantities: the mutual information $MI(A,B)$ and conditional mutual information $CMI(A,B|C)$ as \cite{Ryu:2006bv,Ryu:2006ef}
\be\label{defmi1}
MI(A,B)=S(A)+S(B)-S(AB),
\ee
\be\label{defmi2}
CMI(A,B|C)=S(AC)+S(BC)-S(ABC)-S(C).
\ee
It is believed that mutual information and conditional mutual information are better quantities to measure the entanglement between different subsystems of a thermal system than the entanglement entropy. However, these two quantities are just the linear combination of entanglement entropy but not really new entanglement quantities to describe the entanglement in a thermal state. In recent years a new entanglement quantity based on the purification of thermal state which is called the entanglement of purification (Ep) has been studied from different aspects \cite{Takayanagi:2017knl,Nguyen:2017yqw}. For a thermal state on the boundary time slice, choose two unintersected subsystems $A$ and $B$ the thermal state $\r_{AB}$ can be purified as
\be
\r_{AB}=Tr_{A^{*}B^{*}}(|\sqrt{\r}\rangle\langle\sqrt{\r}|),
\ee
where $|\sqrt{\r}\rangle\langle\sqrt{\r}|=\r$ is a pure state density matrix. Then the entanglement of purification for $A$ and $B$ can be defined as \cite{Barbaran:2002unn}
\be\label{eop1}
Ep(A,B)=\min\limits_{\r_{AB}=Tr_{A^{*}B^{*}}(|\sqrt{\r}\rangle\langle\sqrt{\r}|)}S(\r_{AA^{*}}),
\ee
with $\r_{AA^{*}=Tr_{BB^{*}}(|\sqrt{\r}\rangle\langle\sqrt{\r}|)}$ and $S(\r_{AA^{*}})$ the entanglement entropy associated with $\r_{AA^{*}}$. It is difficult to find the right purification for a general $\r_{AB}$ in the field theory side. The holographic duality of Ep is believed to be the entanglement wedge cross section \cite{Takayanagi:2017knl}
\be\label{ew1}
Ep(A,B)=Ew(A,B)=\frac{Area(\S^{min}_{AB})}{4G_N},
\ee
where $\S^{min}_{AB}$ is the minimal area surface in the entanglement wedge of $A$ and $B$ that ends on their R-T surface as shown in Fig~\ref{entangofpur}: the blue regions are subregions $A$ and $B$ and the red surfaces are the R-T surfaces of $A\cup B$, then the green surface is the minimal surface $\S^{min}_{AB}$.

The remainder of this paper is arranged as follows. In section 2, we give the model setting used in this work. In section 3, we firstly compute the holographic entanglement entropy analytically and derive the minimal area equations. Then we analyze the numerical results of holographic entanglement entropy and compare it with the black hole entropy. We investigate the behavior of other entanglement quantities: mutual information, conditional mutual information and entanglement of purification on the phase diagram in section 4. At last we give the conclusion and discussion in section 5.

\section{Holographic QCD model}

\begin{figure}[t]
	\centerline{\includegraphics[width=10cm]{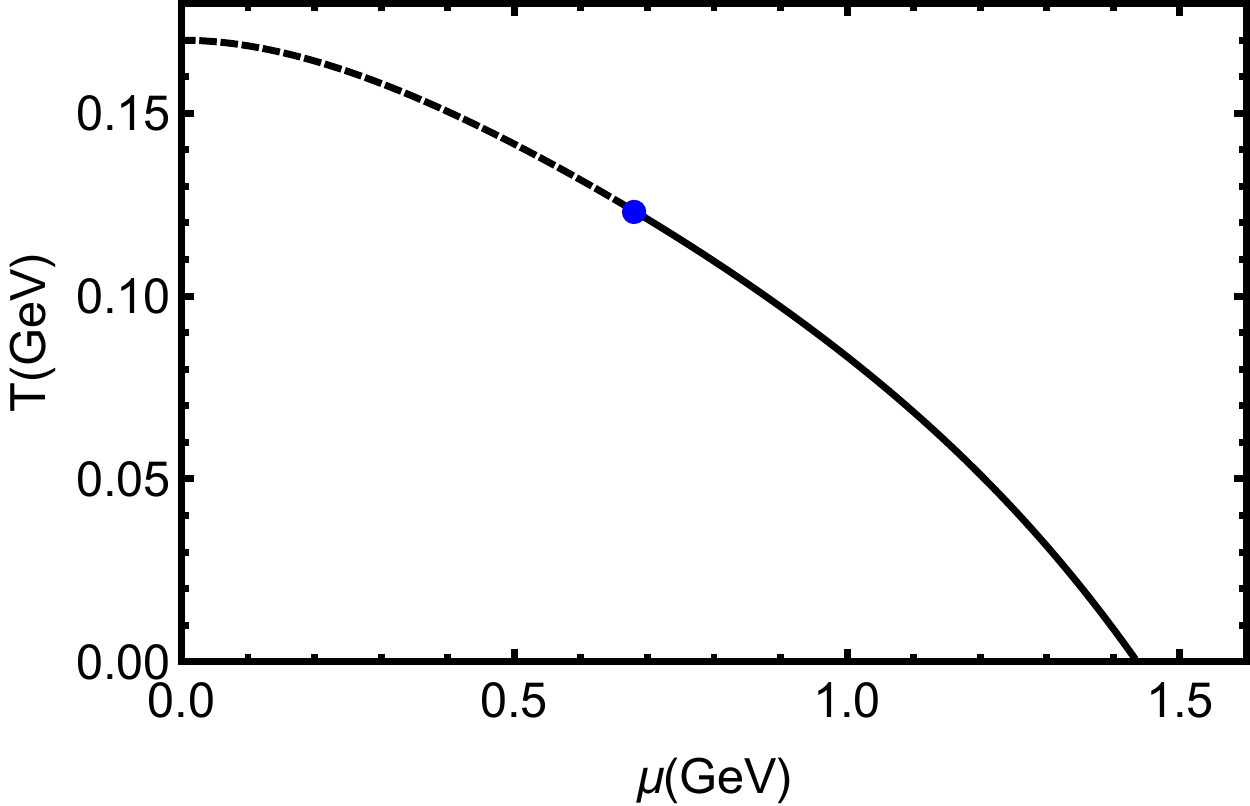}}
	\caption{The phase diagram of the holographic QCD model we used in this work \cite{Yang:2014bqa}. The dashed line is the phase boundary of crossover region ($\m \textless 0.693 \text{GeV}$), the blue dot is the CEP located at ($\m^E=0.693 \text{GeV}$, $T^E=0.121 \text{GeV}$) and the solid line is the first order phase boundary ($\m \textgreater 0.693 \text{GeV}$).}
	\label{phase}
\end{figure}

The holographic QCD model we consider in this work is a 5-dimensional Einstein-Maxwell-dilaton holographic model with the action \cite{Yang:2014bqa}
\be
S=\f{1}{2\k^2} \int d^5 x \sr{-g}\left[R-\f{f\left(\phi\right)}{4}F_{\m\n}^2-\f{1}{2}(\p\phi)^2-V(\phi)\right],
\ee
where $\k^2$ is the gravitational constant and $\k^2=8\pi G_N$. And $g$ is the determinant of the 5-dimensional metric $g_{\a\b}$. The first term $R$ in the action is the Ricci scalar which corresponds to the QCD vacuum sector, the scalar field $\phi$ corresponds to the gluon scalar condensate, and $F_{\m\n}:=\partial_{\m}A_{\n}-\partial_{\n}A_{\m}$ is the strength tensor of  a $U(1)$ gauge field $A_{\m}$, which gives the quark chemical potential and density. Then we take the ansatz of the asymptotic AdS$_5$ metric as \cite{Yang:2014bqa}
\be
ds^2=\f{e^{2A_e(z)}}{z^2}[-\chi(z)dt^2+\f{1}{\chi(z)}dz^2+d\vec{x}^2],
\ee
where $z$ is the holographic direction of asymptotic AdS$_5$, and $z=0$ corresponds to the ultra-violate (UV) boundary spacetime where the QCD theory lives on. Following \cite{Yang:2014bqa}, the dilaton field and the gauge field take the forms of
\be
\phi\equiv \phi(z), \hspace{0.2cm} A_\m dx^\m\equiv A_t(z) dt.
\ee
Then by using the regular boundary conditions at the horizon $z=z_H$ and the asymptotic $AdS_5$ conditions at the UV boundary $z=0$ \cite{Yang:2014bqa}
\be
A_t(z_H)=\chi(z_H)=0,
\ee
\be
A(0)=-\sr{\f{1}{6}}\phi(0), \hspace{0.3cm} \chi(0)=1,
\ee
\be
A_t(0)=\f 1 3\m+3 \r z^2+\cdots,
\ee
with $\m$ and $\r$ the baryon chemical potential and density, respectively,
the warped factor and gauge kinetic function can be fixed as \cite{Yang:2014bqa}
\be
A_e(z)=-\frac{c}{3} z^2-b z^4,
\ee
\be
f(\phi(z))=e^{c z^2-A_e(z)}.
\ee
By solving the equation of motion we get \cite{Yang:2014bqa}
\bea
&& \hspace{-1cm} \chi(z)=1+\frac{1}{\int_0^{z_H} e^{-3 A_e(y)} \, dy}\left\{ \frac{2 c {\mu} ^2}{\left(1-e^{c {z_H}^2}\right)^2} \left[  \int_0^{z_H} e^{-3 A_e(y)} \, dy \int_{z_H}^z e^{c y^2-3 A_e(y)} \, dy \right.\right. \nn\\
&& \hspace{0.2cm}\left. \left.-\int_0^{z_H} e^{c y^2-3 A_e(y)} \, dy \int_{z_H}^z e^{-3 A_e(y)} \, dy\right] -\int_0^z e^{-3 A_e(y)} \, dy\right\} .    \nn\\
\eea
We can also calculate the baryon density $\r$ and the temperature $T$ as \cite{Yang:2014bqa}
\be
\r=\frac{c \m }{9(1-e^{c z_H^2})},
\ee
\bea
&& T=\frac{z_H^3 e^{-3 A_e(z_H)}}{4 \pi  \int_0^{z_H} y^3 e^{-3 A_e(y)} \, dy}\left[1          \right. \nn\\
&& \hspace{0.2cm}\left. -\frac{2 c \mu ^2 \left(e^{c z_H^2} \int_0^{z_H} y^3 e^{-3 A_e(y)} \, dy-\int_0^{z_H} y^3 e^{c y^2-3 A_e(y)} \, dy\right)}{9\left(1-e^{c z_H^2}\right)^2}\right].
\eea
Then by fitting the vacuum vector meson mass $m_{\rho}=0.77 {\rm GeV}$, and the phase transition temperature $T_c=0.17{\rm GeV}$ at $\m=0$ , we can fix $c$ and $b$ as in \cite{Yang:2014bqa}
\be
b=-6.25\times 10^{-4} {\rm GeV}^4, \hspace{3mm} c=0.227 {\rm GeV}^2.
\ee
With above parameters, the phase diagram for deconfinement phase transition of the holographic QCD model of \cite{Yang:2014bqa} is shown in Fig.\ref{phase}. 
The dashed line is the phase boundary of crossover region ($\m \textless 0.693 \text{GeV}$), the blue dot is the CEP located at($\m^E=0.693 \text{GeV}$, $T^E=0.121 \text{GeV}$) and the solid line is the first order phase boundary ($\m \textgreater 0.693 \text{GeV}$).
The black hole entropy of this system can be derived as:
\be
S_{bh}=\f{2\pi}{\k^2} \f{e^{3A_e(z_H)}}{z_H^3}\int_{-\inf}^{\inf}dx_1\int_{-\inf}^{\inf}dx_2\int_{-\inf}^{\inf}dx_3.
\ee
It is obviously that this black hole entropy is divergent. However, because we just need the relation of black hole entropy between different temperature $T$ and baryon chemical potential $\m$ we can just remove the divergent part and define the entropy density as
\be
s_{bh}=\f{2\pi}{\k^2} \f{e^{3A_e(z_H)}}{z_H^3}.
\ee

\section{Holographic entanglement entropy}

In this section, we will consider the holographic entanglement entropy in the holographic QCD model defined in last section. In this work we choose the subregion $A$ to be highly symmetric: 1), a spherical shaped region on the boundary time slice with $0\leq \vec{x}^2\leq r^2_0$ as shown in Fig~\ref{region1}; 2), a strip shaped region on the boundary time slice with $-a/2\leq x_1\leq a/2$ and $-\inf<x_i<\inf$ for $i=2,3$ as shown in Fig~\ref{region2}. The blue region is the region $A$ and the red region is surface $\g_A$ which is the minimal surface that homologous to $A$ in the bulk. It will be very convenient to transform to the spherical coordinate for a spherical shaped region $A$ where the region $A$ will be $0<r<r_0$ as shown by the blue line in Fig~\ref{region1} and the red line is the minimal surface $\g_A$.

\subsection{Minimal area equation}
\begin{figure}[t]
\centerline{\includegraphics[width=11cm]{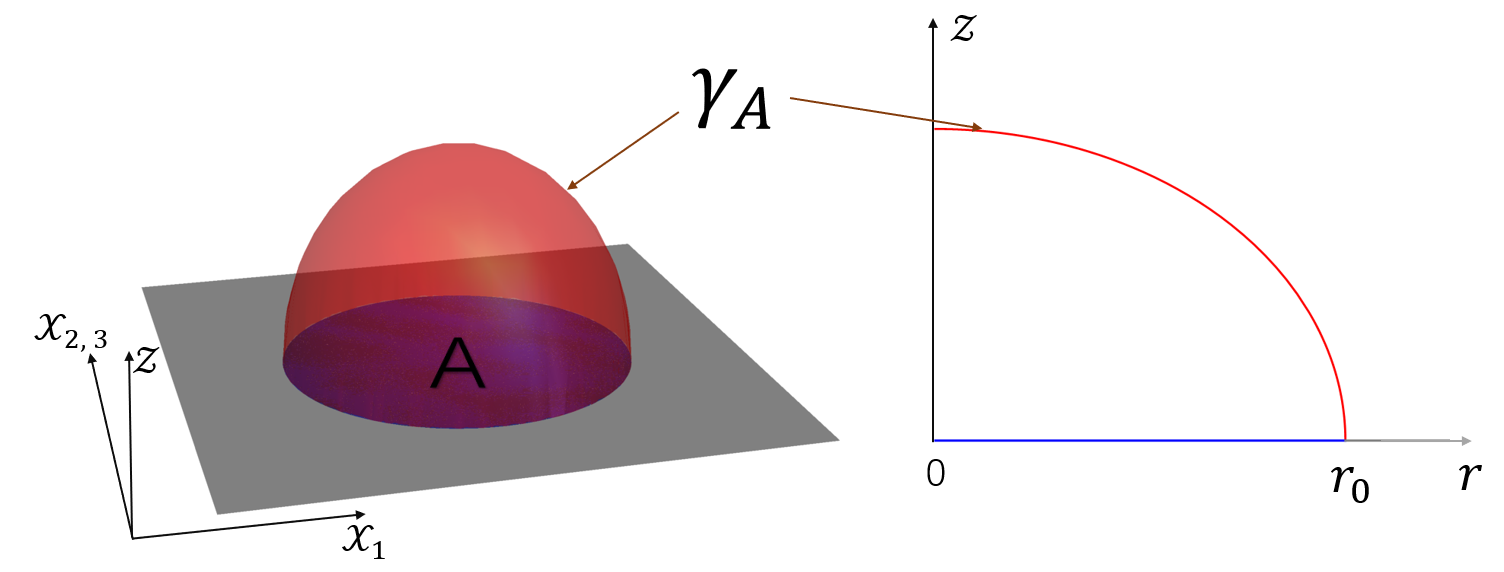}}
\caption{(Left) The spherical shaped subregion $A$ (the blue region) and the minimal surface $\g_A$ (the red surface). (Right) In the spherical coordinate we only need to consider the radius direction and region $A$ is from $r=0$ to $r=r_0$ and the minimal surface $\g_A$ can be determined by $z=z(r)$ which is the solution of the minimal area equation.}
  \label{region1}
\end{figure}
For a spherical shaped region $A$ shown in Fig.\ref{region1} the area of $m(A)$ with $m(A): z=z(x_1, x_2, x_3)$ is
\begin{eqnarray}
 \hspace{-8.5mm}Area(m(A))&=&\int_A \sqrt{h} dx_1 dx_2 dx_3 \nonumber \\
  &=& \int_A \f{e^{3A_e(z)}}{z^3}\sqrt{1+\f{(\partial_{x_1} z)^2+(\partial_{x_2} z)^2+(\partial_{x_3} z)^2}{\chi(z)}}dx_1 dx_2 dx_3.
\end{eqnarray}
After transforming to the spherical coordinate we have $z=z(r, \theta, \phi)$, then use the symmetry of region $A$ and the bulk time slice we know that minimal surface should have the same symmetry with $A$ which means we only need to consider surfaces determined by $z=z(r)$ as shown in Fig~\ref{region1}. Then we have
\bea
Area(m(A))&=&\int_0^{r_0}r^2 dr \int_0^{\pi}\sin(\theta) d\theta \int_0^{2\pi}d\phi \f{e^{3A_e(z)}}{z^3}\sqrt{1+\f{(\partial_{r} z)^2}{\chi(z)}} \nn\\
& =&4\pi \int_0^{r_0}r^2 dr  \f{e^{3A_e(z)}}{z^3}\sqrt{1+\f{(\partial_{r} z)^2}{\chi(z)}}.
\eea
The minimal area equation can be calculated as
\be\label{minibal}
\partial_r^2 z-\frac{2 \left[\chi(z) +\left(\partial_r z\right)^2\right] \left[3 r \chi(z)  \left(z \partial_z A_e(z)-1\right)-2 z \partial_r z\right]+r z \partial_z\chi(z) \left(\partial_r z\right)^2}{2 r z \chi(z) }=0,
\ee
with the boundary conditions
\be
z(r_0)=0, \hspace{0,2cm} \partial_r z(r)|_{r=0}=0,
\ee
we can solve the minimal area equation with the solution $z=z_m(r)$. Then the area of minimal surface is
\be
Area(\g_A)=4\pi \int_0^{r_0}r^2 dr \f{e^{3A_e(z_m)}}{z_m^3}\sqrt{1+\f{(\partial_{r} z_m)^2}{\chi(z_m)}},
\ee
and the holographic entanglement entropy for a spherical shaped region is
\be\label{heebal}
S_A^{sp}=\f{2\pi}{\k^2}Area(\g_A)=\f{8\pi^2}{\k^2}\int_0^{r_0}r^2 dr \f{e^{3A_e(z_m)}}{z_m^3}\sqrt{1+\f{(\partial_{r} z_m)^2}{\chi(z_m)}}.
\ee

\begin{figure}[t]
  \centerline{\includegraphics[width=11cm,keepaspectratio]{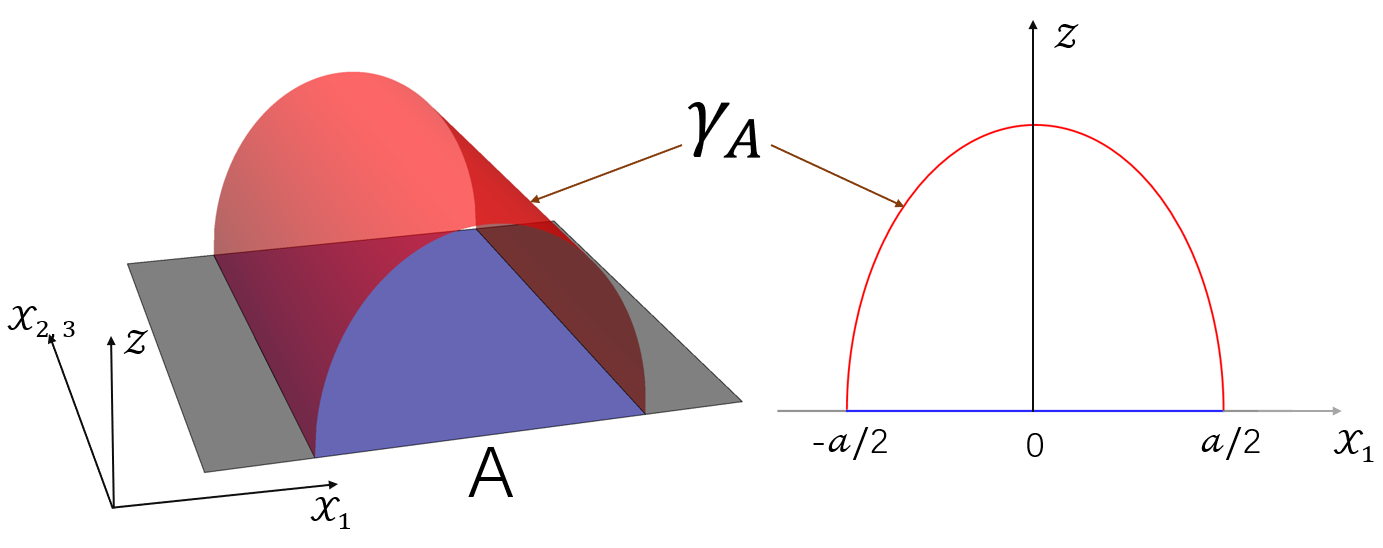}}
  \caption{(Left) The strip shaped subregion $A$(the blue region) and the minimal surface $\g_A$(the red surface). (Right) Because of the symmetry of region $A$ we only need to consider the $x_1$ direction and the region $A$ is $-a/2<x_1<a/2$ (the blue line) and the minimal surface $\g_A$ is determined by $z=z(x_1)$ which is the solution of the minimal area equation.}
  \label{region2}
\end{figure}
For a strip shaped region $A$ shown in Fig.\ref{region2}, the area of surface $m(A)$ is
\be
 Area(m(A))=\int_A \sqrt{h} dx_1 dx_2 dx_3=\int_A \f{e^{3A_e(z)}}{z^3}\sqrt{1+\f{(\partial_{x_1} z)^2+(\partial_{x_2} z)^2+(\partial_{x_3} z)^2}{\chi(z)}}dx_1 dx_2 dx_3.
\ee
Also use the symmetry of region $A$ and the bulk time slice we know that the minimal surface should have the same symmetry with $A$ and can be determined by $z=z(x_1)$ as shown in Fig~\ref{region2}. Then the area of $m(A)$
\be
 Area(m(A))=\int_A \f{e^{3A_e(z)}}{z^3}\sqrt{1+\f{(\partial_{x_1} z)^2}{\chi(z)}}dx_1 dx_2 dx_3 = M_1 M_2\int_{-\f a 2}^{\f a 2} \f{e^{3A_e(z)}}{z^3}\sqrt{1+\f{(\partial_{x_1} z)^2}{\chi(z)}}dx_1,
\ee
with $M_1$, $M_2$ the length of region $A$ along the $x_2$ and $x_3$ direction. Then the minimal area equation shows
\be\label{ministr}
\partial_{x_1}^2 z-\frac{3(z \partial_z A_e(z)-1)\left[\chi(z)+(\partial_{x_1}z)^2\right]}{z }-\frac{\partial_z\chi(z)(\partial_{x_1}z)^2}{2\chi(z)}=0.
\ee
Then we can solve this minimal area equation with the boundary conditions
\be
z(a/2)=0, \hspace{0.2cm}  z(-a/2)=0,
\ee
or equally
\be
z(a/2)=0, \hspace{0.2cm}  \partial_{x_1}z(x_1)|_{x_1=0}=0.
\ee
Plug the solution of minimal area equation $z=z_m(x_1)$ into area formula we get the area of minimal surface $\g_A$
\be
Area(\g_A)=2 M_1 M_2\int_0^{\f a 2} \f{e^{3A_e(z_m)}}{z_m^3}\sqrt{1+\f{(\partial_{x_1} z_m)^2}{\chi(z_m)}}dx_1.
\ee
And the holographic entanglement entropy for a strip shaped region is
\be\label{heestr}
S_A^{st}=\f{2\pi }{\k^2}Area(\g_A)=\f{4\pi M_1 M_2}{\k^2}\int_0^{\f a 2} \f{e^{3A_e(z_m)}}{z_m^3}\sqrt{1+\f{(\partial_{x_1} z_m)^2}{\chi(z_m)}}dx_1.
\ee

\subsection{Numerical results}

In this part, we will firstly give the numerical results of the black hole entropy of the holographic QCD model, then show the numerical results for the holographic entanglement entropy of a spherical shaped region and a strip shaped region.

\subsubsection{The black hole entropy}\label{bhd}

\begin{figure}[t]
    \centerline{\includegraphics[width=13cm,keepaspectratio]{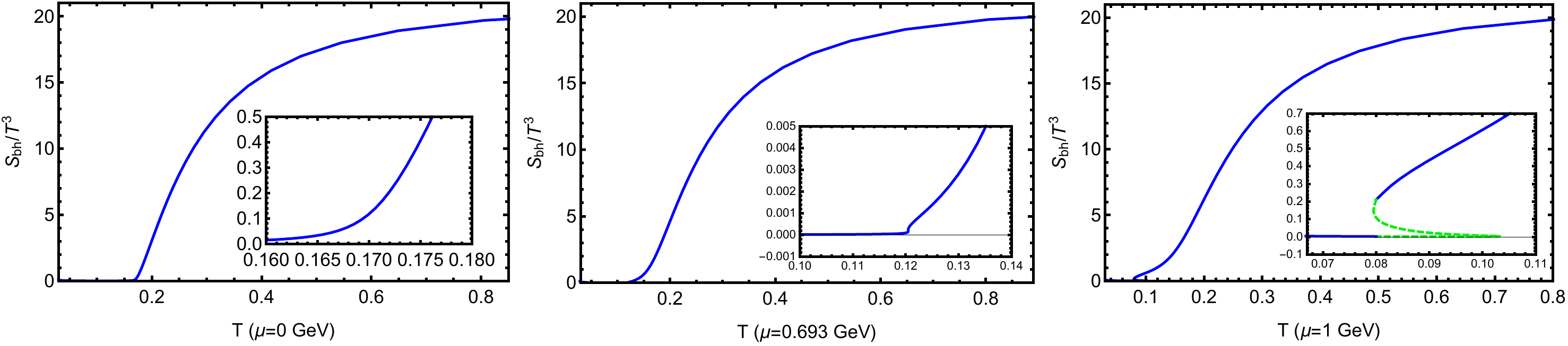}}
  \caption{The scaled black hole entropy $S_{bh}/T^3$ as a function of the temperature for different quark chemical potentials. The blue solid lines are the physical black hole entropy from the minimal of free energy. At $\m=0$ the phase transition happens at $T=0.17~{GeV}$, and it is a crossover so the $S_{bh}$ is single valued and smooth. When $\m=0.693~\rm{GeV}$ it is the critical end point of the first order phase transition and the transition temperature is $T=0.121~\rm{GeV}$ where the $S_{bh}$ is single valued but not smooth. With $\m=1~\rm{GeV}$ the first order phase transition happens at $T=0.08~\rm{GeV}$ where the $S_{bh}$ is not single valued and the physical value of $S_{bh}$ is not continued at this point. }\label{sbh}
\end{figure}

Before we show the entanglement entropy on the QCD phase diagram, we firstly calculate the black hole entropy for different temperature $T$ and baryon chemical potential $\m$.  First of all we fix the constant $\k^2=1$ in our following calculation. And here we choose the divergent term to be
\be
\int_{-\inf}^{\inf}dx_1\int_{-\inf}^{\inf}dx_2\int_{-\inf}^{\inf}dx_3 \rightarrow 0.11 \rm{GeV}^{-3},
\ee
and note that we can fix this term to be any constant in principle. Then the black hole entropy $S_{bh}$ for fixed $\m$ as a function of the temperature is shown in Fig~\ref{sbh}.

The blue solid line is the physical value of $S_{bh}/T^3$, and for first order phase transition the green dashed line is not physical corresponding to the metastable state determined by the maximum of the free energy \cite{Yang:2014bqa}. It is noticed that at different chemical potential, the black hole entropy $S_{bh}$ or equally $S_{bh}/T^3$ is almost zero in the hadron phase and then sharply increases at the phase boundary. And at very high temperature the ratio of $S_{bh}/T^3$ goes to a constant about $20$ for all temperatures. Then focus on the behavior of the black hole entropy at the phase boundary we find that in the crossover region ($0\leq \m \textless 0.693$ $\rm{GeV}$), the ratio of $S_{bh}/T^3$ is single valued and smooth. At the CEP ($\m^E=0.693~\rm{GeV}$, $T^E=0.121~\rm{GeV}$), the ratio of $S_{bh}/T^3$ is single valued but not smooth. And in the first order phase transition region ($0.693~\rm{GeV} \textless\m$), the ratio of $S_{bh}/T^3$ is not single valued which means that the ratio of $S_{bh}/T^3$ is not continued at the first order phase boundary. Beyond all doubt that the black hole entropy or the normalized black hole entropy is just the holographic duality of the entropy of thermal QCD.

\subsubsection{The holographic entanglement entropy } \label{sub_hee}

In the following we will perform the numerical calculation of the holographic entanglement entropy for a spherical shaped region and a strip shaped region.
Note that to get a finite area of the minimal surface we also need to choose a UV-cutoff because of the boundary UV-divergence at $z=0 ~\rm{GeV}^{-1}$ which is called the renormalization of holographic entanglement entropy \cite{Ryu:2006bv,Ryu:2006ef}.

\begin{figure}[t]
  \centerline{ \includegraphics[width=11cm,keepaspectratio]{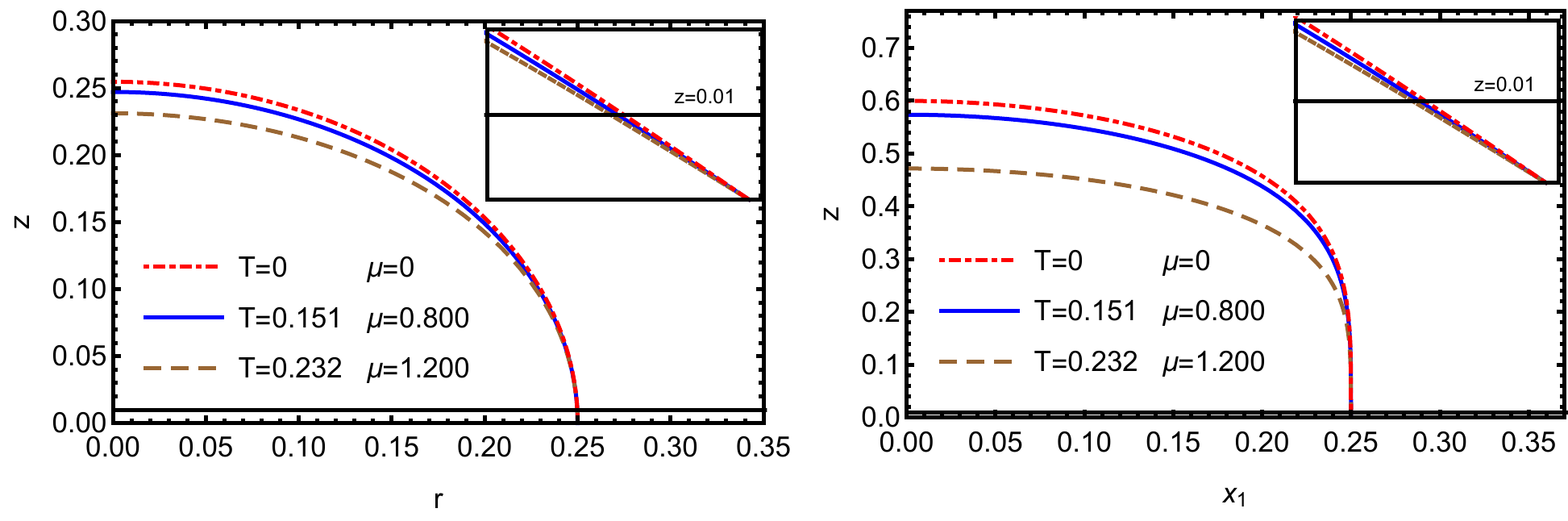}}
  \caption{The minimal surface with a UV-cutoff at $z=0.01$ $\rm{GeV}^{-1}$ with different temperatures and baryon chemical potentials. (Left) The minimal surface for a spherical shaped region $A$. (Right) The minimal surface for a strip shaped region $A$. Here the unit for the temperature and the chemical potential $\mu$ is in ${\rm GeV}$. }
  \label{reg1}
\end{figure}

\begin{figure}[t]

   \centerline{ \includegraphics[width=11cm,keepaspectratio]{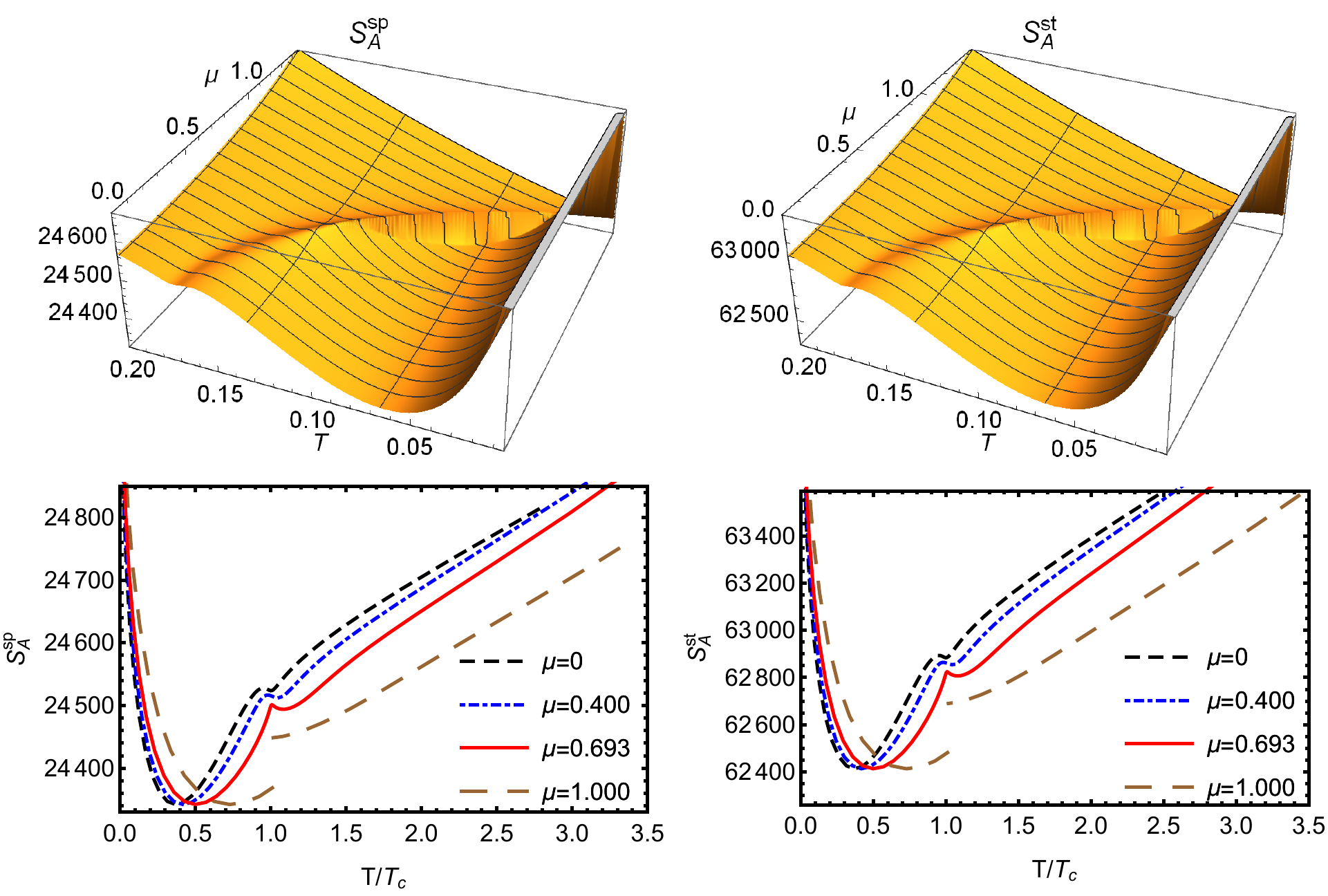}}
  \caption{3D-plot of the holographic entanglement entropy on $(T,\mu)$ plane and the 2D-plot along fixed baryon chemical potential line. Here the unit for the temperature $T$ and the chemical potential $\mu$ is in ${\rm GeV}$.}\label{3dsbss}
\end{figure}

 Firstly, we take the UV-cutoff to be $z=\e$ and the area of minimal surface for a spherical shaped region and a strip shaped region are
\be\label{eeball}
S_A^{sp}=\f{8\pi^2}{\k^2}\int_0^{r_0-\e_0}r^2 dr \f{e^{3A_e(z_m)}}{z_m^3}\sqrt{1+\f{(\partial_{r} z_m)^2}{\chi(z_m)}},
\ee
\be\label{eestrip}
S_A^{st}=\f{2\pi }{\k^2}Area(\g_A)=\f{4\pi M_1 M_2}{\k^2} \int_0^{\f a 2-\e_0} \f{e^{3A_e(z_m)}}{z_m^3}\sqrt{1+\f{(\partial_{x_1} z_m)^2}{\chi(z_m)}}dx_1.
\ee
We choose $\e=0.01~\rm{GeV}^{-1}$ in our calculation and so $\e_0$ is not a constant but a function of the temperature $T$ and the quark chemical potential $\m$ e.g. $\e_0=\e_0(T, \m)$ as shown in Fig~\ref{reg1}.
Every minimal surface for different $T$ and $\m$ ends at the same point on the boundary with $z=0 $ $\rm{GeV}^{-1}$. After taking the UV-cutoff, different surfaces have different boundaries on the slice of $z=0.01 ~\rm{GeV}^{-1}$. For the strip shaped region $A$, we also need to take the parameters $M_1$ and $M_2$ to be finite, and we choose $M_1=M_2=1$ and $r_0=a/2=0.25~\rm{GeV}^{-1}$. Then within the holographic QCD model and using the holographic entanglement entropy formulae Eqs. (\ref{eeball}) and (\ref{eestrip}), we can calculate the holographic entanglement entropy between the two subregions $A$ and $\bar{A}$ for different temperatures $T$ and baryon chemical potentials $\m$. Fig~\ref{3dsbss} shows the 3D-plot of the holographic entanglement entropy on $(T,\mu)$ plane and the 2D-plot along fixed baryon chemical potential line. Note that the physical entanglement entropy could be determined from the minimal value of the free energy.

From Fig~\ref{3dsbss} it is clear that the holographic entanglement entropy of a spherical shaped region $S_A^{sp}$ and of a strip shaped region $S_A^{st}$ are very similar on the $(T,\mu)$ phase diagram. In the crossover region both $S_A^{sp}$ and $S_A^{st}$ decrease firstly at low temperature and then increase but then decrease again, thus form a weak peak structure around the phase boundary, and then increase sequentially in the QGP phase. But note that although $S_A^{sp}$ and $S_A^{st}$ have very similar increasing and decreasing behavior on the phase diagram they have totally different value for a fixed $T$ and $\m$. Near the phase boundary $S_A^{sp}$ and $S_A^{st}$ change smoothly in the crossover region and form a weak peak structure around the phase boundary. However, we find that the top of the peak is not exactly the phase boundary. At the CEP ($\m^E=0.693$ $\rm{GeV}$ $T^E=0.121$ $\rm{GeV}$) $S_A^{sp}$ and $S_A^{st}$ are continued but not smooth. And note that the CEP is exactly at the top of the peak around the phase boundary. In the first order phase transition region $S_A^{sp}$ and $S_A^{st}$ are not continued at the phase boundary. These continuity properties are very similar to the case of black hole entropy $S_{bh}$ and so the holographic entanglement entropy of different boundary regions could also be a signal of QCD phase transition.

\subsubsection{High temperature behavior}

\begin{figure}[t]
  \centerline{  \includegraphics[width=11cm,keepaspectratio]{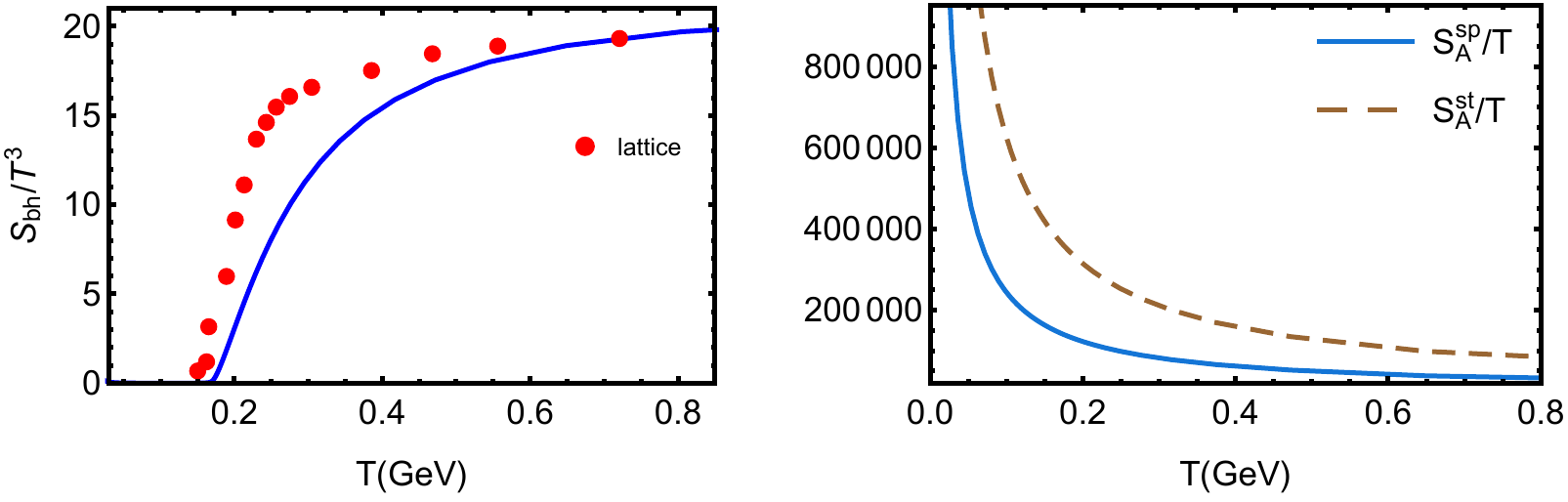}}
  \caption{The behavior of black hole entropy and entanglement entropy at high temperature with $\m=0$. (Left) The red dots are the lattice data \cite{Karsch:2007dp} of thermal entropy and the blue solid line is the black hole entropy in our holographic QCD model. (Right) $S_A^{sp}/T$ and $S_A^{st}/T$ at $\m=0$.}\label{Fighight}
\end{figure}

In section \ref{bhd} we concluded that the black hole entropy is the holographic duality of thermal entropy of the QCD. Lattice results \cite{Karsch:2007dp} show that the thermal entropy $S_{th}$ has the behavior $S_{th}\sim T^3$ at high temperature with $\m=0$. It is obviously the black hole entropy $S_{bh}$ has the same property but the holographic entanglement entropy does not. At high temperature the behavior of holographic entanglement entropy are more likely proportional to $T$ with $\m=0$ as shown in Fig~\ref{Fighight}.

Black hole entropy $S_{bh}/T^3$ shown as blue solid line in the left figure matches the lattice result of thermal entropy $S_{th}/T^3$ and they both go to a constant. The holographic entanglement entropy for both regions are proportional to $T$ at high temperature when $\m=0$.

\section{Other entanglement properties}

As we know that entanglement entropy measures how strong the entanglement between different subsystems is when the whole system is in a pure state. However, for a thermal state system the entanglement entropy contain the contributions of the thermodynamics but not only the entanglement contributions. From the inequalities of entanglement entropy like the subadditivity and the strong subadditivity one can define other entanglement quantities. The most often encountered two are called the mutual information $MI(A,B)$ (\ref{defmi1}) and the conditional mutual information $CMI(A,B|C)$  (\ref{defmi2}) \cite{Ryu:2006bv,Ryu:2006ef}. Another entanglement quantity that is helpful to study the entanglement property in a thermal system is the entanglement of purification (\ref{eop1}) and its holographic duality the entanglement wedge cross section (\ref{ew1}) \cite{Takayanagi:2017knl,Nguyen:2017yqw}.

In this section we only consider strip shaped region $A$ and $B$ defined as $a\leq x_1\leq b$ and $-\inf<x_i<\inf$ and once we fixed their boundaries $a$ and $b$ the regions will be determined. The three situations of $A$ and $B$ we considered are shown in Table~\ref{coeffi}.

 \begin{table}\centering
 	\begin{tabular}{|c|c|c|c|c|}\hline

 \multirow{2}{*}{\diagbox [width=7em,trim=l] {Case} {Interval}}
 & \multicolumn{2}{|c|}{A}         & \multicolumn{2}{|c|}{B} \\ \cline{2-5}
  		&  a      &     b      &    a       &      b                    \\ \hline

 		1                &   -0.5      &     0.2      &    0.3       &      0.5         \\ \hline
 		2                &   -0.5      &     0.1      &    0.2       &      0.5         \\ \hline
 		3                &   -0.5      &     -0.05    &    0.05      &      0.5         \\ \hline
 	\end{tabular}
\caption{Three situations of the two subregions $A$ and $B$ we used in this section. Note that for each subregion we just need to fix the two boundaries $a$ and $b$ that are shown in the second and third column (subregion $A$) or forth and fifth column (subregion $B$) in the third to fifth row of Table~\ref{coeffi}.}  \label{coeffi}
 \end{table}

\subsection{Mutual information}\label{MI}

\begin{figure}[t]
  \centerline{\includegraphics[width=11cm,keepaspectratio]{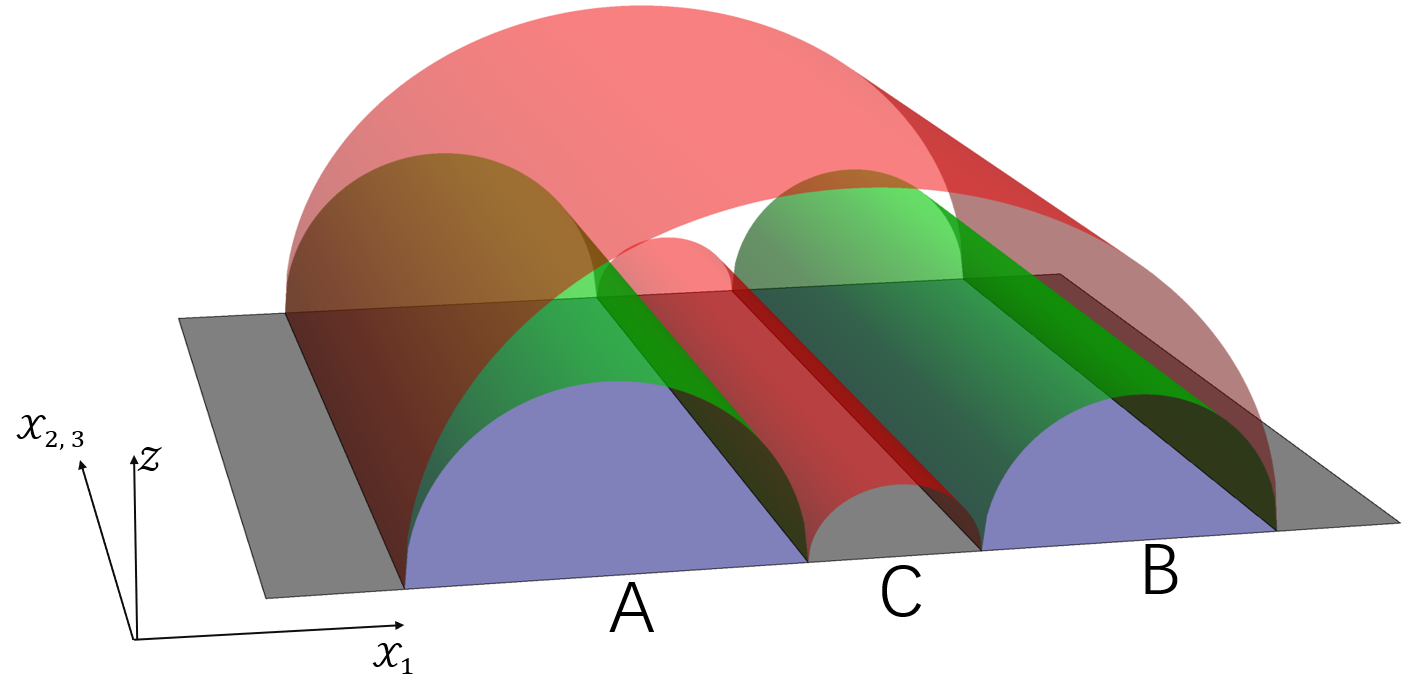}}
	\caption{The mutual information. The green surfaces are the R-T surfaces of $A$ and $B$ and the red surfaces are the R-T surface of $A\cup B$.}\label{mutual}
\end{figure}

\begin{figure}[t]
  \centerline{\includegraphics[width=11cm,keepaspectratio]{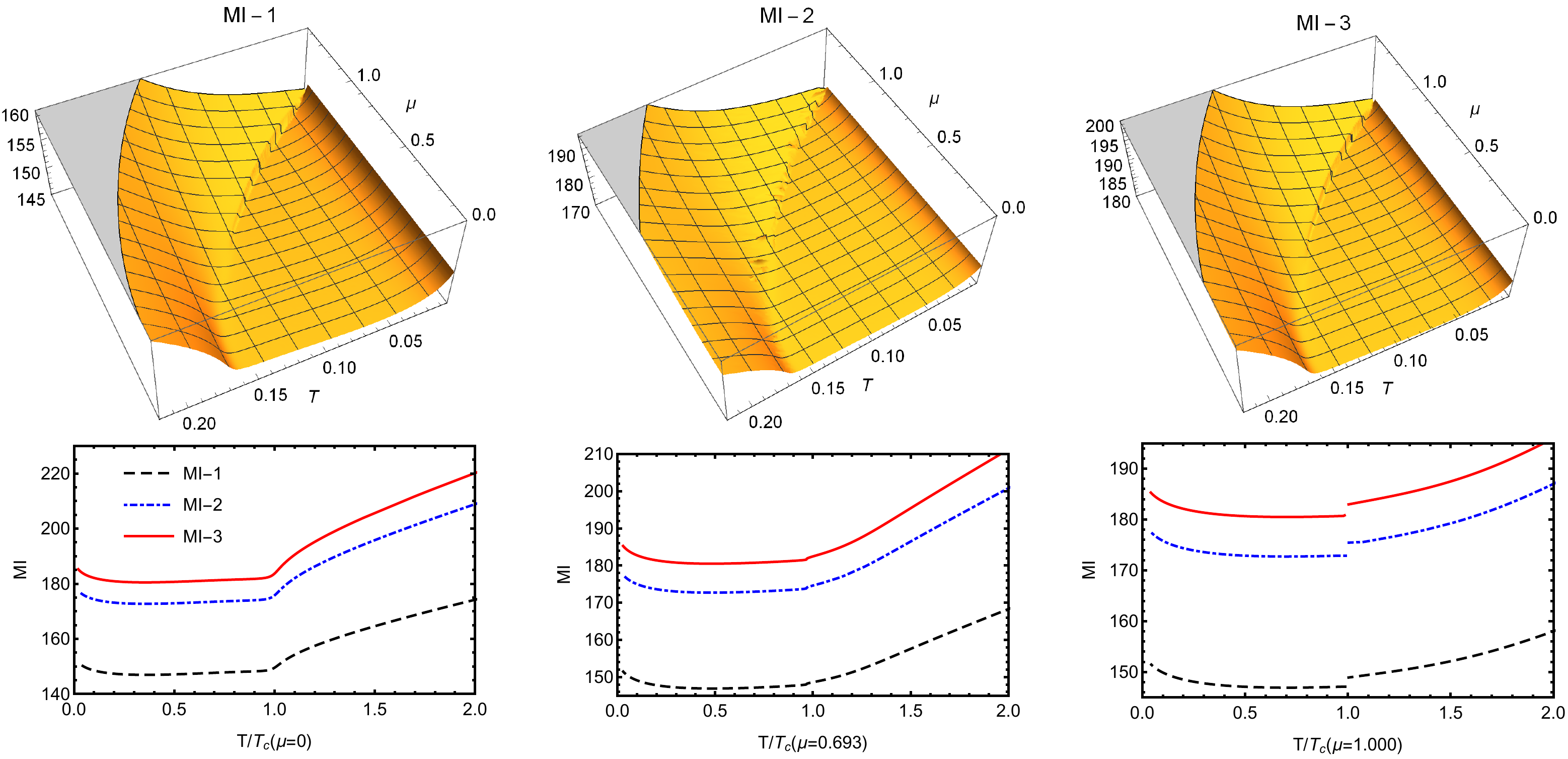}}
	\caption{(Upper) 3D-plot of the mutual information on $(T,\mu)$ plane and (Lower) the 2D-plot along fixed baryon chemical potential line for case-1,2,3 of $A$ and $B$ in Table~\ref{coeffi}. Here the unit for the temperature $T$ and the chemical potential $\mu$ is in ${\rm GeV}$.}\label{mulre}
\end{figure}

In this section we consider the mutual information of two unintersected subsystems $A$ and $B$ as shown in Fig~\ref{mutual}.
The mutual information $MI(A,B)$ is defined as
\be
MI(A,B)=S(A)+S(B)-S(AB).
\ee
And we choose three nontrivial constructions of $A$ and $B$ as shown in Table~\ref{coeffi} which makes $MI(A,B)> 0$ for any temperature $T$ and baryon chemical potential $\m$, then the holographic mutual information can be calculated as
\be
MI(A,B)=S(A)+S(B)-S(AB) =\frac{1}{4G_N}[Area(\S_{gre})-Area(\S_{red})],
\ee
where $Area(\S_{gre})$ ($Area(\S_{red})$) means the area of green (red) surfaces as shown in Fig~\ref{mutual}. Note that the green surfaces are the R-T surfaces of $A$ and $B$, respectively. And the red surfaces correspond to the R-T surfaces of $A\cup B$. The numerical results of mutual information for different setting of $A$ and $B$ are shown in Fig~\ref{mulre}.

Here we denote the mutual information $MI(A,B)$ of the three cases of $A$ and $B$ we considered in Table~\ref{coeffi} as MI-1, MI-2 and MI-3 correspondingly. The upper sub-figure of Fig~\ref{mulre} shows the 3D-plot of mutual information between $A$ and $B$ on the ($T$, $\m$) phase diagram. It is clear that $MI(A,B)$ for different constructions of $A$ and $B$  behave similarly on the phase diagram if ignore there exact value but have different behavior with the entanglement entropy as shown in Fig~\ref{3dsbss}. $MI(A,B)$ do not change a lot in the hadron matter phase and then increase with the increase of $T$ and $\m$ in the QGP phase. Just like the entanglement entropy, in the crossover region $MI(A,B)$ is single valued and change smoothly near the phase boundary, is continuous but not smooth at the CEP and is not single valued (and so is not continuous) at the first order phase boundary. Not surprisingly, $MI(A,B)$ is finite at any $T$ and $\m$ although $S(A)$, $S(B)$ and $S(AB)$ are divergent. The lower 2D-plot in Fig~\ref{mulre} shows the same results. While the 2D-plot gives another interesting result
\be
\text{MI-1}\leq \text{MI-2}\leq \text{MI-3}.
\ee
In consideration of the symmetry of the bulk spacetime we have $Area(\S_{red}^1)=Area(\S_{red}^2)=Area(\S_{red}^3)$. Here the superscript means the three cases of $A$ and $B$, respectively. Then this inequality means $S(A)+S(B)$ decreases with the increase of $|Area(A)-Area(B)|$ when $Area(A)+Area(B)$ is a constant. Where $Area(A)$ and $Area(B)$ means the area of subregion $A$ and $B$, respectively. When $Area(A)=Area(B)$ the $S(A)+S(B)$ will take the maximal value.

\subsection{Conditional mutual information}
\begin{figure}[t]
  \centerline{\includegraphics[width=11cm,keepaspectratio]{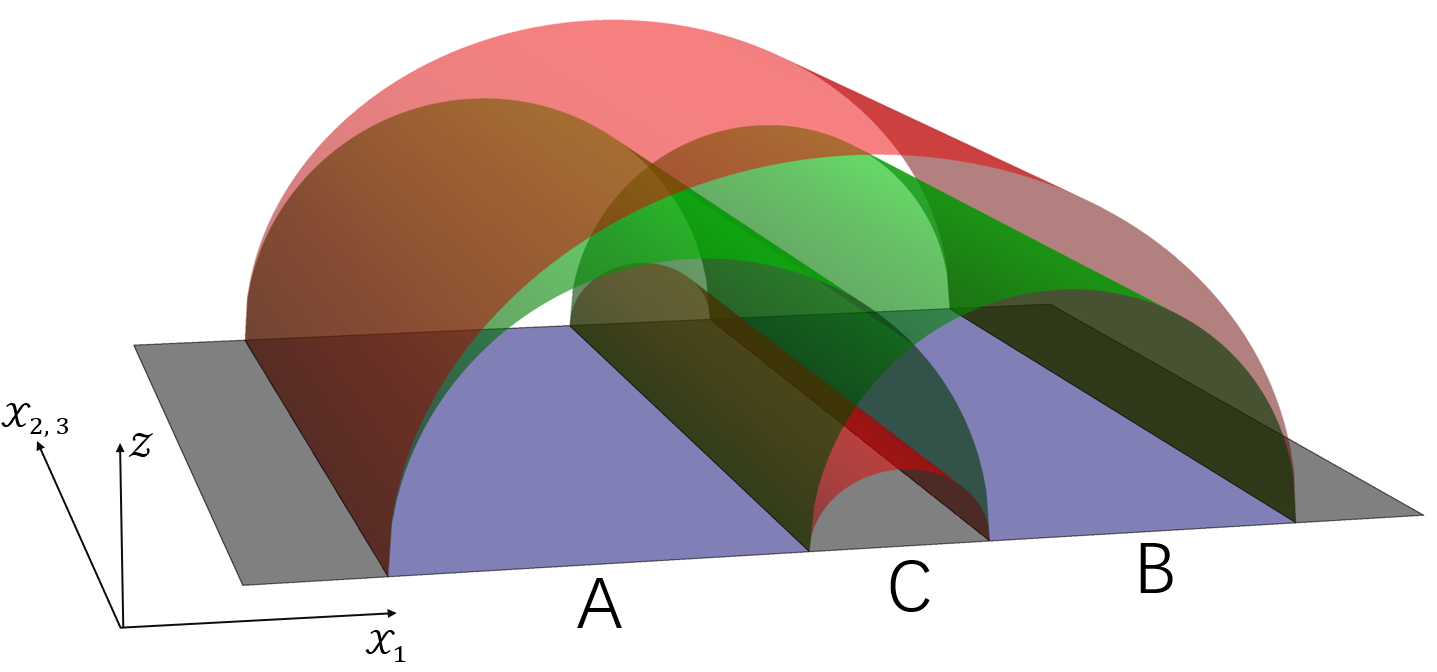}}
	\caption{The conditional mutual information. The green surfaces are the R-T surfaces of $A\cup C$ and $B\cup C$ and the red surfaces are the R-T surfaces of $C$ and $A\cup B\cup C$.}\label{condition}
\end{figure}

In this section we consider the conditional mutual information of two unintersected subsystems $A$ and $B$ as shown in Fig~\ref{condition}.

The conditional mutual information
\be
CMI(A,B|C)=S(AC)+S(BC)-S(ABC)-S(C).
\ee
Then the holographic mutual information can be calculated as
\be
CMI(A,B|C)=S(AC)+S(BC)-S(ABC)-S(C) =\frac{1}{4G_N}[Area(\S_{gre})-Area(\S_{red})]
\ee
where $Area(\S_{gre})$ ($Area(\S_{red})$) means the area of green (red) surfaces as shown in Fig~\ref{condition}. Here the green surfaces are the R-T surfaces of $A\cup C$ and $B\cup C$, respectively. And the red surfaces correspond to the R-T surfaces of $A\cup B\cup C$ and $C$, respectively.

\begin{figure}[t]
	  \centerline{
	\includegraphics[width=11cm,keepaspectratio]{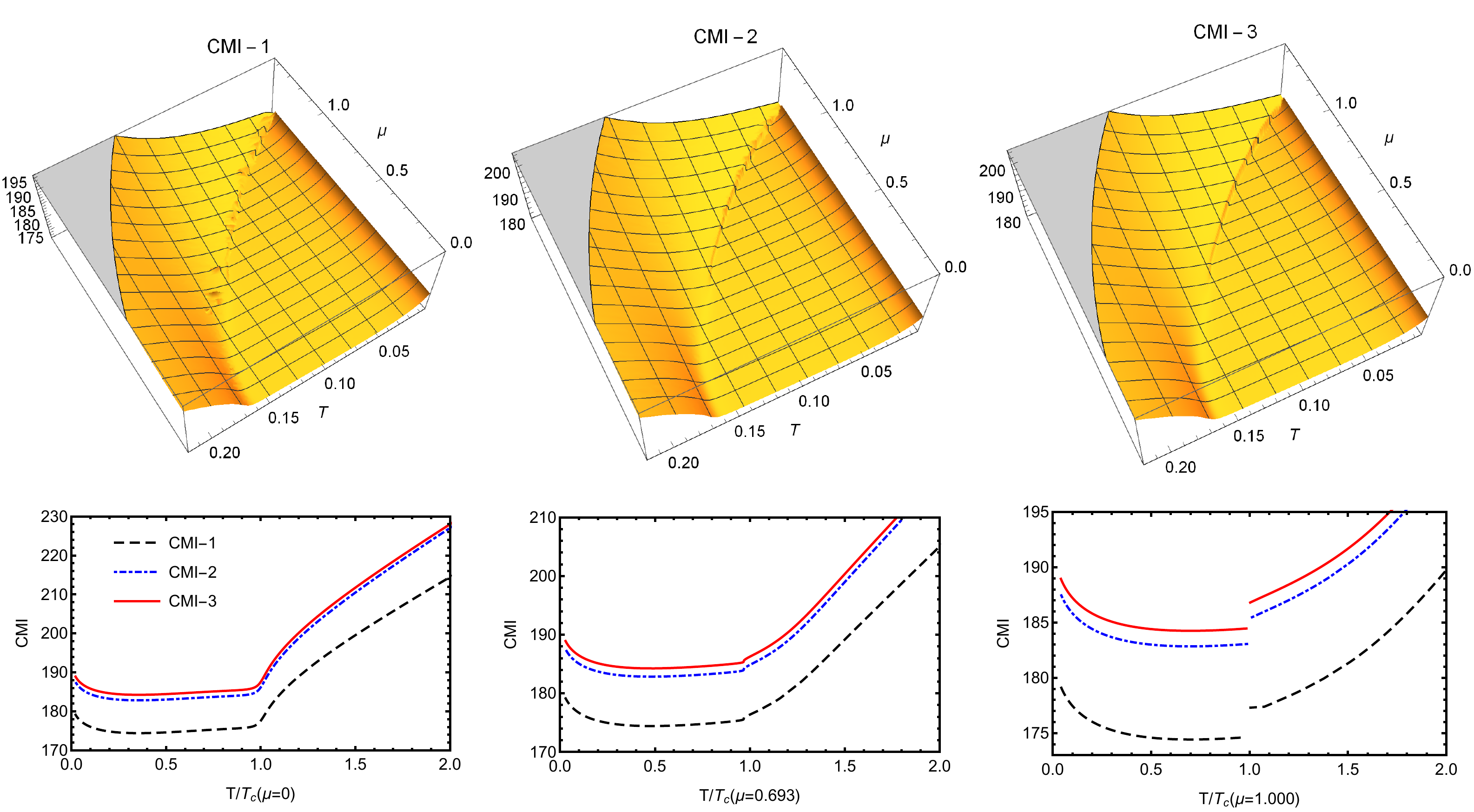}}
	\caption{(Upper) 3D-plot of the conditional mutual information on $(T,\mu)$ plane and (Lower) the 2D-plot along fixed baryon chemical potential line for case-1,2,3 of $A$ and $B$ in Table~\ref{coeffi}. Here the unit for the temperature $T$ and the chemical potential $\mu$ is in ${\rm GeV}$.}\label{conmulre}
\end{figure}

For the conditional mutual information we also consider the same regions $A$ and $B$  as given in Table~\ref{coeffi}. And Fig~\ref{conmulre} shows the 3D-plot of $CMI(A,B|C)$ on the ($T$, $\m$) phase diagram and its 2D-plot along fixed $\m$.
Note that just like the mutual information case we also denote the three settings of $A$ and $B$ as CMI-1, CMI-2 and CMI-3. For different $A$, $B$ and $C$, $CMI(A,B|C)$ also behave similarly on the ($T$, $\m$) phase diagram but with different values. What is more is that the behavior of $CMI(A,B|C)$ is very similar to $MI(A,B)$, and it does not change so much in the hadronic matter phase and then rise up quickly with the increase of $T$ and $\m$ in the QGP phase. Near the phase boundary, $CMI(A,B|C)$ changes smoothly in the crossover region, continuously but not smoothly at CEP and shows discontinuity behavior in the first phase transition region. Which means $CMI(A,B|C)$ could also measure the entanglement between different subregion $A$ and $B$ and denote the phase transition of strongly coupled matter. The lower 2D-plot in Fig~\ref{conmulre} exhibits the same results and also gives similar inequality
\be
\text{CMI-1}\leq \text{CMI-2}\leq \text{CMI-3}.
\ee
Following the same discussion in section \ref{MI}: firstly consider the symmetry of the bulk spacetime we have $Area(\S_{red}^1)=Area(\S_{red}^2)=Area(\S_{red}^3)$. Here the superscript also means the three cases of $A$ and $B$, respectively. Then this inequality means $S(AC)+S(BC)$ decreases with the increase of $|Area(A\cup C)-Area(B\cup C)|$ when $Area(A\cup C)+Area(B\cup C)$ is a constant. With $Area(A\cup C)$ and $Area(B\cup C)$ correspond to the area of subregion $A\cup C$ and $B\cup C$, respectively.  And when $Area(A\cup C)=Area(B\cup C)$ the $S(AC)+S(BC)$ will take the maximal value.

\subsection{Entanglement of purification}

\begin{figure}[t]
	  \centerline{\includegraphics[width=11cm,keepaspectratio]{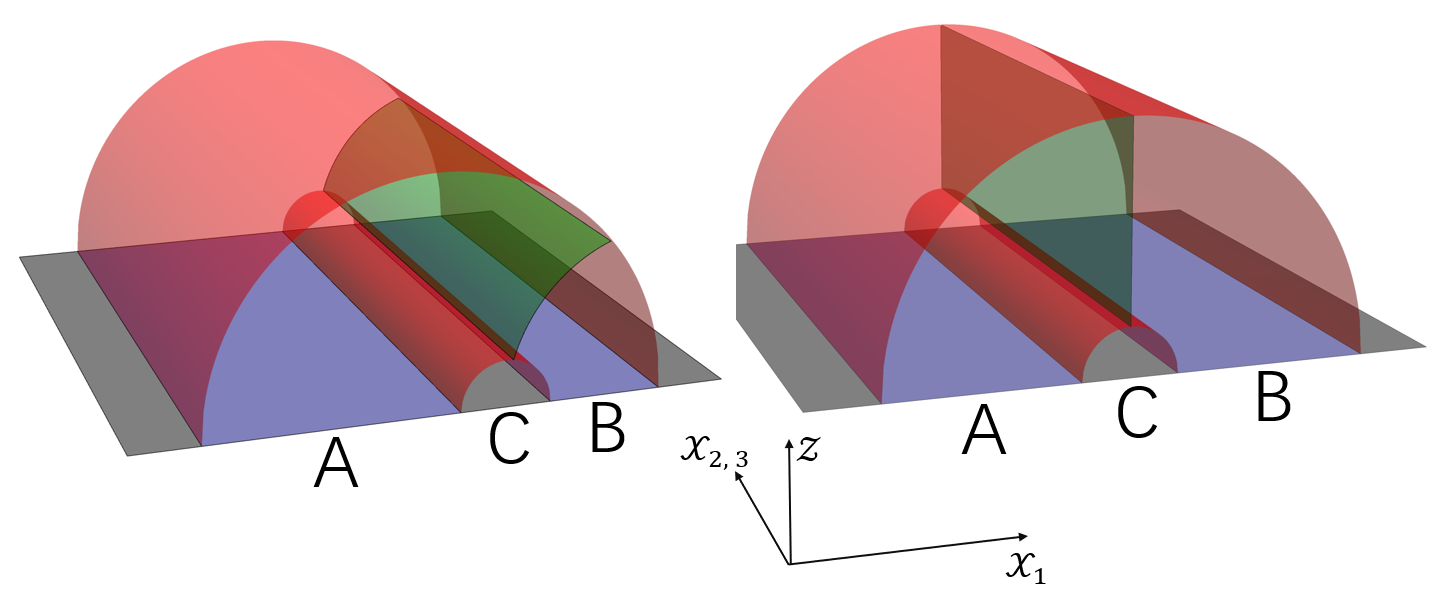}}
	\caption{The entanglement of purification. (Left) The asymmetry case and (Right) The symmetry case of $A$ and $B$. The red surfaces are the R-T surface of $A\cup B$ and the green surface gives the minimal cross section of the entanglement wedge of $A$ and $B$. }\label{entangofpur}
\end{figure}

\begin{figure}[t]
	  \centerline{\includegraphics[width=11cm]{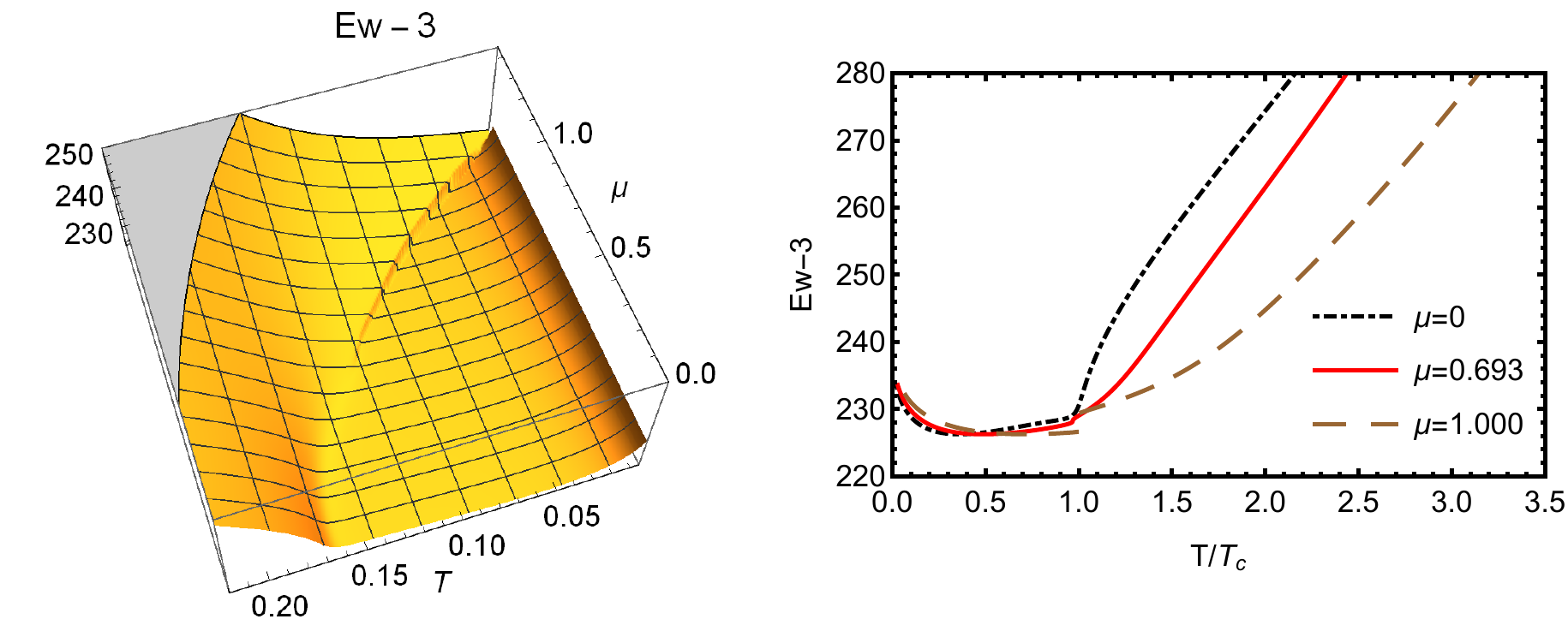}}
	\caption{(Left) 3D-plot of the entanglement of purification on $(T,\mu)$ plane and (Right) the 2D-plot along fixed baryon chemical potential line for case-3 of $A$ and $B$ in Table~\ref{coeffi}. Here the unit for the temperature $T$ and the chemical potential $\mu$ is in ${\rm GeV}$.}\label{entwed3}
\end{figure}

\begin{figure}[t]
	  \centerline{\includegraphics[width=11cm,keepaspectratio]{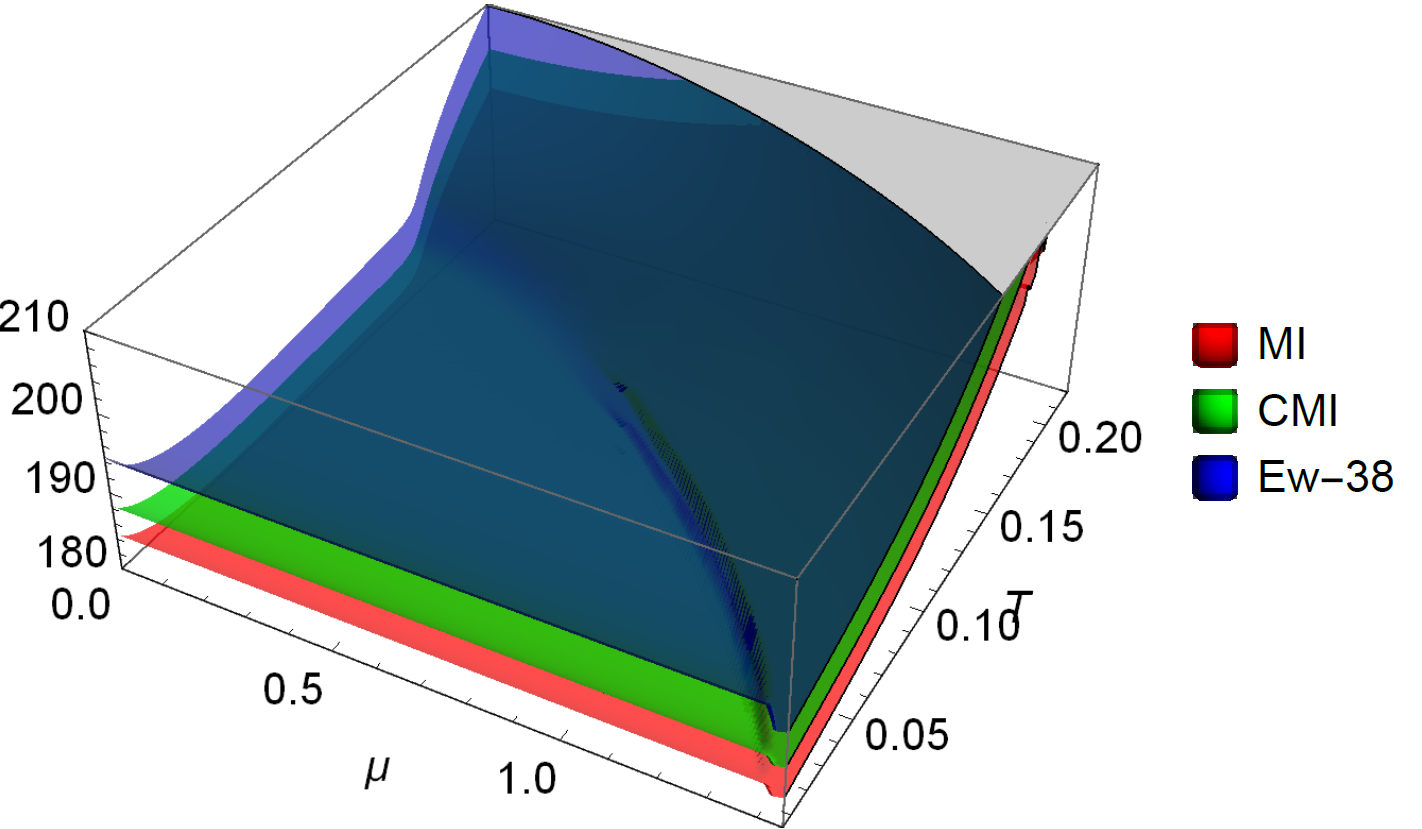}}
	\caption{The mutual information(MI), conditional mutual information(CMI) and entanglement of purification(Ew) for case-3 of $A$ and $B$ in Table~\ref{coeffi}.}\label{compar3}
\end{figure}

In this section we consider the entanglement of purification of two unintersected subsystems $A$ and $B$ as shown in Fig~\ref{entangofpur}. The holographic duality of entanglement of purification is the entanglement wedge cross section \cite{Takayanagi:2017knl,Nguyen:2017yqw}
\be
Ep(A,B)=Ew(A,B)=\frac{Area(\S^{min}_{AB})}{4G_N}=\frac{Area(\S_{gre})}{4G_N},
\ee
where $Area(\S_{gre})$ means the area of green surfaces as shown in Fig~\ref{entangofpur}.
In this section we only consider the symmetric case of $A$ and $B$ (case-3 in Table~\ref{coeffi}). The minimal surface is just the surface with $x_1=const$ as shown the right sub-figure in Fig~\ref{entangofpur}. The asymmetric case of $A$ and $B$ could be much more complected and we may consider it in our next work. Fig~\ref{entwed3} shows the 3D-plot of $Ep(A,B)$ on the ($T$, $\m$) phase diagram and its 2D-plot along fixed $\m$.
In \cite{Freedman:2016zud} the authors propose that the $Ew(A,B)$ could equals to the mutual information by introducing a group of bit flow related to the mutual information which can be limited within the entanglement wedge and ends on $A$ and $B$. And in \cite{Takayanagi:2017knl,Nguyen:2017yqw} the authors suggested $Ep(A,B)$ should be the holographic duality of the entanglement of purification of $A$ and $B$. The 3D-plot in Fig~\ref{entwed3} gives very similar behavior of $Ep(A,B)$ to $MI(A,B)$ and $CMI(A,B|C)$ as shown in Fig~\ref{mulre} and Fig~\ref{conmulre}. These similarities could also be seen from Fig~\ref{compar3} where we plot the $MI(A,B)$ (the red plot), $CMI(A,B|C)$ (the green plot) and $Ep(A,B)-38$ (the blue plot). Note that to compare these three entanglement quantities we only consider the symmetric case of  $A$ and $B$ (case-3 in Table~\ref{coeffi}). It is clear in Fig~\ref{compar3} that for fixed $T$ and $\m$ $Ep(A,B)\geq CMI(A,B|C)\geq MI(A,B)$. The second "$\geq$" is because the monogamy of mutual information. If we consider the mutual information of $A$, $B$ and $C$ then we have
\bea
\hspace{-7mm} I(A,B,C)&=&S(A)+S(B)+S(C)-S(AB)-S(BC)-S(AC)+S(ABC) \nn\\
&=&-I(A,B|C)+I(A,B)\leq 0. 
\eea
And the first "$\geq$" suggest that the maximal number of allowed bit threads that connect $A$ and $B$ does not equal to the $MI(A,B)$ or $CMI(A,B|C)$.

\section{Conclusion and discussion}

In this work we study the holographic entanglement entropy of a holographic QCD model with a critical end point. We consider the behavior of entanglement entropy for a strip shaped region and a spherical shaped region on the phase diagram. And find that the behavior of holographic entanglement entropy on the phase diagram is independent with the shape of region $A$ expect the exact value although the minimal area equations for different $A$ are totally different. Then we study how other entanglement quantities include the mutual information, conditional mutual information and the entanglement of purification behave on the phase diagram. We find that the three entanglement quantities have very similar behavior:  their values do not change so much in the hadronic matter phase and then rise up quickly with the increase of $T$ and $\m$ in the QGP phase. Near the phase boundary, these three entanglement quantities change smoothly in the crossover region, continuously but not smoothly at CEP and show discontinuity behavior in the first phase transition region. At last we find an inequality for $Ep(A,B)$, $I(A,B|C)$ and $I(A,B)$
\be
Ep(A,B)\geq CMI(A,B|C)\geq MI(A,B),
\ee
at any $T$ and $\m$. This inequality suggest that the monogamy of $I(A,B,C)$ is still satisfied and $Ep(A,B)$ is not the holographic duality of mutual information or conditional mutual information.

It should be noted that the black hole entropy for different holographic QCD model have very similar behavior, however, even for $\m=0$ the behavior of holographic entanglement entropy does depend on details of models. For different models the behavior of holographic entanglement entropy on the ($T$, $\m$) phase diagram could be totally different. Which means the geometry in the bulk spacetime is totally different. However, in principle the entanglement entropy of different subsystems of QCD should be unique. So this model dependence of holographic entanglement entropy means we need to find a way to fix the bulk geometry. One way could be the machine learning \cite{Hashimoto:2018ftp,Hashimoto:2018bnb} and another is to build a full dynamic holographic QCD model that we are still working on.

\section*{Acknowledgements}

We would like to thank Peng Liu for very helpful discussion. Z.L. acknowledge support by the NSFC under Grant No. 11947233, CPSF under Grant No. 2019M662507 and the start-up funding from Zhengzhou University. The work of M.H. was supported in part by the NSFC under Grant Nos. 11725523, 11735007, 11261130311 (CRC 110 by DFG and NSFC), Chinese Academy of Sciences under Grant No. XDPB09, and the start-up funding from University of Chinese Academy of Sciences.

\providecommand{\href}[2]{#2}\begingroup\raggedright\endgroup


\begin{thebibliography}{10}
	
	\bibitem{VonNeumann1932}
	V.~Neumann, {\em Mathematische Grundlagen der Quantenmechanik}.
	\newblock Springer, Berlin, German, 1932.
	
	\bibitem{Maldacena:1997re}
	J.~M. Maldacena, ``{The Large N limit of superconformal field theories and
		supergravity},'' \href{http://dx.doi.org/10.1023/A:1026654312961}{{\em Int.
			J. Theor. Phys.} {\bfseries 38} (1999) 1113--1133},
	\href{http://arxiv.org/abs/hep-th/9711200}{{\ttfamily arXiv:hep-th/9711200
			[hep-th]}}.
	[Adv. Theor. Math. Phys. {\bf 2}, (1998) 231].
	
	\bibitem{Gubser:1998bc}
	S.~Gubser, I.~R. Klebanov, and A.~M. Polyakov, ``{Gauge theory correlators from
		noncritical string theory},''
	\href{http://dx.doi.org/10.1016/S0370-2693(98)00377-3}{{\em Phys. Lett.}
		{\bfseries B428} (1998) 105--114},
	\href{http://arxiv.org/abs/hep-th/9802109}{{\ttfamily arXiv:hep-th/9802109
			[hep-th]}}.
	
	\bibitem{Witten:1998qj}
	E.~Witten, ``{Anti-de Sitter space and holography},'' {\em Adv. Theor. Math.
		Phys.} {\bfseries 2} (1998) 253--291,
	\href{http://arxiv.org/abs/hep-th/9802150}{{\ttfamily arXiv:hep-th/9802150
			[hep-th]}}.
	
	\bibitem{Aharony:1999ti}
	O.~Aharony, S.~S. Gubser, J.~M. Maldacena, H.~Ooguri, and Y.~Oz, ``{Large N
		field theories, string theory and gravity},''
	\href{http://dx.doi.org/10.1016/S0370-1573(99)00083-6}{{\em Phys.Rept.}
		{\bfseries 323} (2000) 183--386},
	\href{http://arxiv.org/abs/hep-th/9905111}{{\ttfamily arXiv:hep-th/9905111
			[hep-th]}}.
	
	\bibitem{VanRaamsdonk:2010pw}
	M.~Van~Raamsdonk, ``{Building up spacetime with quantum entanglement},''
	\href{http://dx.doi.org/10.1007/s10714-010-1034-0,
		10.1142/S0218271810018529}{{\em Gen. Rel. Grav.} {\bfseries 42} (2010)
		2323--2329}, \href{http://arxiv.org/abs/1005.3035}{{\ttfamily arXiv:1005.3035
			[hep-th]}}.
	[Int. J. Mod. Phys.D19,2429(2010)].
	
	\bibitem{Faulkner:2013ica}
	T.~Faulkner, M.~Guica, T.~Hartman, R.~C. Myers, and M.~Van~Raamsdonk,
	``{Gravitation from Entanglement in Holographic CFTs},''
	\href{http://dx.doi.org/10.1007/JHEP03(2014)051}{{\em JHEP} {\bfseries 03}
		(2014) 051},
	\href{http://arxiv.org/abs/1312.7856}{{\ttfamily arXiv:1312.7856 [hep-th]}}.
	
	\bibitem{Swingle:2014uza}
	B.~Swingle and M.~Van~Raamsdonk, ``{Universality of Gravity from
		Entanglement},''
	\href{http://arxiv.org/abs/1405.2933}{{\ttfamily arXiv:1405.2933 [hep-th]}}.
	
	\bibitem{Freedman:2016zud}
	M.~Freedman and M.~Headrick, ``{Bit threads and holographic entanglement},''
	\href{http://dx.doi.org/10.1007/s00220-016-2796-3}{{\em Commun. Math. Phys.}
		{\bfseries 352} no.~1, (2017) 407--438},
	\href{http://arxiv.org/abs/1604.00354}{{\ttfamily arXiv:1604.00354 [hep-th]}}.
	
	\bibitem{Czech:2017ryf}
	B.~Czech, ``{Einstein Equations from Varying Complexity},''
	\href{http://dx.doi.org/10.1103/PhysRevLett.120.031601}{{\em Phys. Rev.
			Lett.} {\bfseries 120} no.~3, (2018) 031601},
	\href{http://arxiv.org/abs/1706.00965}{{\ttfamily arXiv:1706.00965 [hep-th]}}.
	
	\bibitem{Erdmenger:2007cm}
	J.~Erdmenger, N.~Evans, I.~Kirsch, and E.~Threlfall, ``{Mesons in Gauge/Gravity
		Duals - A Review},'' \href{http://dx.doi.org/10.1140/epja/i2007-10540-1}{{\em
			Eur. Phys. J.} {\bfseries A35} (2008) 81--133},
	\href{http://arxiv.org/abs/0711.4467}{{\ttfamily arXiv:0711.4467 [hep-th]}}.
	
	\bibitem{deTeramond:2012rt}
	G.~F. de~Teramond and S.~J. Brodsky, ``{Hadronic Form Factor Models and
		Spectroscopy Within the Gauge/Gravity Correspondence},'' in {\em
		{Proceedings, Ferrara International School Niccolò Cabeo: Hadron
			Electromagnetic Form Factors: Ferrara, Italy, May 23-28, 2011}}, pp.~54--109.
	\newblock 2011.
	\newblock \href{http://arxiv.org/abs/1203.4025}{{\ttfamily arXiv:1203.4025
			[hep-ph]}}.
	\newblock
	\newblock
	
	\bibitem{Kim:2012ey}
	Y.~Kim, I.~J. Shin, and T.~Tsukioka, ``{Holographic QCD: Past, Present, and
		Future},'' \href{http://dx.doi.org/10.1016/j.ppnp.2012.09.002}{{\em Prog.
			Part. Nucl. Phys.} {\bfseries 68} (2013) 55--112},
	\href{http://arxiv.org/abs/1205.4852}{{\ttfamily arXiv:1205.4852 [hep-ph]}}.
	
	\bibitem{Adams:2012th}
	A.~Adams, L.~D. Carr, T.~Schäfer, P.~Steinberg, and J.~E. Thomas, ``{Strongly
		Correlated Quantum Fluids: Ultracold Quantum Gases, Quantum Chromodynamic
		Plasmas, and Holographic Duality},''
	\href{http://dx.doi.org/10.1088/1367-2630/14/11/115009}{{\em New J. Phys.}
		{\bfseries 14} (2012) 115009},
	\href{http://arxiv.org/abs/1205.5180}{{\ttfamily arXiv:1205.5180 [hep-th]}}.
	
	\bibitem{Ryu:2006bv}
	S.~Ryu and T.~Takayanagi, ``{Holographic derivation of entanglement entropy
		from AdS/CFT},'' \href{http://dx.doi.org/10.1103/PhysRevLett.96.181602}{{\em
			Phys. Rev. Lett.} {\bfseries 96} (2006) 181602},
	\href{http://arxiv.org/abs/hep-th/0603001}{{\ttfamily arXiv:hep-th/0603001
			[hep-th]}}.
	
	\bibitem{Ryu:2006ef}
	S.~Ryu and T.~Takayanagi, ``{Aspects of Holographic Entanglement Entropy},''
	\href{http://dx.doi.org/10.1088/1126-6708/2006/08/045}{{\em JHEP} {\bfseries
			0608} (2006) 045},
	\href{http://arxiv.org/abs/hep-th/0605073}{{\ttfamily arXiv:hep-th/0605073
			[hep-th]}}.
	
	\bibitem{Zhang:2016rcm}
	S.-J. Zhang, ``{Holographic entanglement entropy close to crossover/phase
		transition in strongly coupled systems},''
	\href{http://dx.doi.org/10.1016/j.nuclphysb.2017.01.010}{{\em Nucl. Phys.}
		{\bfseries B916} (2017) 304--319},
	\href{http://arxiv.org/abs/1608.03072}{{\ttfamily arXiv:1608.03072 [hep-th]}}.
	
	\bibitem{Knaute:2017lll}
	J.~Knaute and B.~Kämpfer, ``{Holographic Entanglement Entropy in the QCD Phase
		Diagram with a Critical Point},''
	\href{http://dx.doi.org/10.1103/PhysRevD.96.106003}{{\em Phys. Rev.}
		{\bfseries D96} no.~10, (2017) 106003},
	\href{http://arxiv.org/abs/1706.02647}{{\ttfamily arXiv:1706.02647 [hep-ph]}}.
	
	\bibitem{Ali-Akbari:2017vtb}
	M.~Ali-Akbari and M.~Lezgi, ``{Holographic QCD, entanglement entropy, and
		critical temperature},''
	\href{http://dx.doi.org/10.1103/PhysRevD.96.086014}{{\em Phys. Rev.}
		{\bfseries D96} no.~8, (2017) 086014},
	\href{http://arxiv.org/abs/1706.04335}{{\ttfamily arXiv:1706.04335 [hep-th]}}.
	
	\bibitem{Dong:2016wcf}
	X.~Dong, ``{Shape Dependence of Holographic Rényi Entropy in Conformal Field
		Theories},'' \href{http://dx.doi.org/10.1103/PhysRevLett.116.251602}{{\em
			Phys. Rev. Lett.} {\bfseries 116} no.~25, (2016) 251602},
	\href{http://arxiv.org/abs/1602.08493}{{\ttfamily arXiv:1602.08493 [hep-th]}}.
	
	\bibitem{Bianchi:2016xvf}
	L.~Bianchi, S.~Chapman, X.~Dong, D.~A. Galante, M.~Meineri, and R.~C. Myers,
	``{Shape dependence of holographic Rényi entropy in general dimensions},''
	\href{http://dx.doi.org/10.1007/JHEP11(2016)180}{{\em JHEP} {\bfseries 11}
		(2016) 180},
	\href{http://arxiv.org/abs/1607.07418}{{\ttfamily arXiv:1607.07418 [hep-th]}}.
	
	\bibitem{Cavini:2019wyb}
	G.~Cavini, D.~Seminara, J.~Sisti, and E.~Tonni, ``{On shape dependence of
		holographic entanglement entropy in AdS$_{4}$/CFT$_{3}$ with Lifshitz scaling
		and hyperscaling violation},''
	\href{http://arxiv.org/abs/1907.10030}{{\ttfamily arXiv:1907.10030 [hep-th]}}.
	
	\bibitem{Chen:2017ahf}
	B.~Chen, Z.~Li, and J.-j. Zhang, ``{Corrections to holographic entanglement
		plateau},'' \href{http://dx.doi.org/10.1007/JHEP09(2017)151}{{\em JHEP}
		{\bfseries 09} (2017) 151},
	\href{http://arxiv.org/abs/1707.07354}{{\ttfamily arXiv:1707.07354 [hep-th]}}.
	
	\bibitem{Takayanagi:2017knl}
	T.~Takayanagi and K.~Umemoto, ``{Entanglement of purification through
		holographic duality},''
	\href{http://dx.doi.org/10.1038/s41567-018-0075-2}{{\em Nature Phys.}
		{\bfseries 14} no.~6, (2018) 573--577},
	\href{http://arxiv.org/abs/1708.09393}{{\ttfamily arXiv:1708.09393 [hep-th]}}.
	
	\bibitem{Nguyen:2017yqw}
	P.~Nguyen, T.~Devakul, M.~G. Halbasch, M.~P. Zaletel, and B.~Swingle,
	``{Entanglement of purification: from spin chains to holography},''
	\href{http://dx.doi.org/10.1007/JHEP01(2018)098}{{\em JHEP} {\bfseries 01}
		(2018) 098},
	\href{http://arxiv.org/abs/1709.07424}{{\ttfamily arXiv:1709.07424 [hep-th]}}.
	
	\bibitem{Barbaran:2002unn}
	B.~M. Terhal, M.~Horodecki, D.~W. Leung, and D.~P. DiVincenzo, ``{The
		entanglement of purification},''
	\href{http://dx.doi.org/10.1063/1.1498001}{{\em J. Math. Phys.} {\bfseries
			43} (2002) 4286},
	\href{http://arxiv.org/abs/quant-ph/0202044}{{\ttfamily arXiv:quant-ph/0202044
			[quant-ph]}}.
	
	\bibitem{Yang:2014bqa}
	Y.~Yang and P.-H. Yuan, ``{A Refined Holographic QCD Model and QCD Phase
		Structure},'' \href{http://dx.doi.org/10.1007/JHEP11(2014)149}{{\em JHEP}
		{\bfseries 11} (2014) 149},
	\href{http://arxiv.org/abs/1406.1865}{{\ttfamily arXiv:1406.1865 [hep-th]}}.
	
	\bibitem{Karsch:2007dp}
	F.~Karsch, ``{Recent lattice results on finite temperature and density QCD.
		Part I.},'' \href{http://dx.doi.org/10.22323/1.047.0026}{{\em PoS} {\bfseries
			CPOD07} (2007) 026},
	\href{http://arxiv.org/abs/0711.0656}{{\ttfamily arXiv:0711.0656 [hep-lat]}}.
	
	\bibitem{Hashimoto:2018ftp}
	K.~Hashimoto, S.~Sugishita, A.~Tanaka, and A.~Tomiya, ``{Deep learning and the
		AdS/CFT correspondence},''
	\href{http://dx.doi.org/10.1103/PhysRevD.98.046019}{{\em Phys. Rev.}
		{\bfseries D98} no.~4, (2018) 046019},
	\href{http://arxiv.org/abs/1802.08313}{{\ttfamily arXiv:1802.08313 [hep-th]}}.
	
	\bibitem{Hashimoto:2018bnb}
	K.~Hashimoto, S.~Sugishita, A.~Tanaka, and A.~Tomiya, ``{Deep Learning and
		Holographic QCD},'' \href{http://dx.doi.org/10.1103/PhysRevD.98.106014}{{\em
			Phys. Rev.} {\bfseries D98} no.~10, (2018) 106014},
	\href{http://arxiv.org/abs/1809.10536}{{\ttfamily arXiv:1809.10536 [hep-th]}}.
	
\end{thebibliography}

\end{document}